\documentclass[prd,eqsecnum,twocolumn,amsfonts,showpacs]{revtex4}

\input epsf

\usepackage{graphicx}

\usepackage{bm}

\newcommand{\beq}{\begin{equation}}
\newcommand{\eeq}{\end{equation}}
\newcommand{\beqs}{\begin{eqnarray}}
\newcommand{\eeqs}{\end{eqnarray}}

\newcommand{\gsim}{\mathrel{\raisebox{-.6ex}{$\stackrel{\textstyle>}{\sim}$}}}

\begin{document}

\title{Zeros of the Potts Model Partition Function in the Large-$q$ Limit}

\author{Shu-Chiuan Chang$^{a,b}$ \thanks{email: scchang@mail.ncku.edu.tw} 
Robert Shrock$^{c}$ \thanks{email: robert.shrock@sunysb.edu}}

\bigskip

\affiliation{(a) \ Department of Physics \\
National Cheng Kung University \\
Tainan 70101, Taiwan} 

\bigskip

\affiliation{(b) \ Physics Division \\
National Center for Theoretical Science \\
National Taiwan University \\
Taipei 10617, Taiwan} 

\bigskip

\affiliation{(c) \ C. N. Yang Institute for Theoretical Physics \\
State University of New York \\
Stony Brook, N. Y. 11794 }

\begin{abstract}

We study the zeros of the $q$-state Potts model partition function
$Z(\Lambda,q,v)$ for large $q$, where $v$ is the temperature variable and
$\Lambda$ is a section of a regular $d$-dimensional lattice with coordination
number $\kappa_\Lambda$ and various boundary conditions.  We consider the
simultaneous thermodynamic limit and $q \to \infty$ limit and show that when
these limits are taken appropriately, the zeros lie on the unit circle
$|x_\Lambda|=1$ in the complex $x_\Lambda$ plane, where $x_\Lambda=v
q^{-2/\kappa_\Lambda}$.  For large finite sections of some lattices we also
determine the circular loci near which the zeros lie for large $q$.


\end{abstract}

\maketitle

\newpage
\pagestyle{plain}
\pagenumbering{arabic}

\section{Introduction}
\label{sectionI}

In this paper we shall study the $q$-state Potts model in the limit $q \to
\infty$ and shall present some new results on the complex-temperature zeros of
the partition function in this limit. The Potts model has served as a valuable
model for the study of phase transitions and critical phenomena
\cite{wurev,baxterbook}.  The $q \to \infty$ limit of the model is exactly
solvable on any lattice \cite{pg}, as can be seen from the fact that in this
limit the model essentially reduces to a single-site problem.  Since the $q \to
\infty$ limit of the Potts model is solvable, it is natural to investigate the
properties of the model for large $q$.  Indeed, although we will focus on
partition function zeros rather than physical applications, we note that the
Potts model with large $q$ has been used in modelling the kinetic behavior of
soap froths \cite{gag}.

The (zero-field) Potts model is defined, for temperature $T$ on a lattice
$\Lambda$, or more generally, a graph $G$, by the partition function
\beq
Z(G,q,v) = \sum_{ \{ \sigma_n \} } 
\exp ( K\sum_{\langle i j \rangle} \delta_{\sigma_i \sigma_j} ) 
\label{z}
\eeq
where $\sigma_i=1,...,q$ are the classical spin variables on each vertex (site)
$i \in G$, $\langle i j \rangle$ denotes pairs of adjacent vertices, $K=\beta
J$ where $\beta = (k_BT)^{-1}$, and $J$ is the spin-spin coupling.  We use the
notation $a=e^K$, and $v=a-1=e^K-1$.  For the Potts ferromagnet, $J > 0$ so $v
\ge 0$, while for the antiferromagnet, $J < 0$ so that $-1 \le v \le 0$.  The
graph $G$ is formally defined by its set of vertices $V$ and its set of edges
(= bonds) $E$; we denote the number of vertices of $G$ as $n=n(G)=|V|$ and the
number of edges of $G$ as $e(G)=|E|$.  The complex-temperature zeros of
$Z(G,q,v)$, i.e., the zeros in the complex $v$ plane, have the property that on
a regular lattice graph with dimensionality greater than the lower critical
dimensionality, $d \ge 2$, in the thermodynamic limit $L_j \to \infty$ with
$L_j/L_k$ a finite nonzero constant, $1 \le j,k \le d$, they merge to form a
locus ${\cal B}$ consisting of curves that bound regions of different analytic
behavior for the corresponding free energy $F = -k_BTf$, where the
dimensionless function $f$ is given by $f = \lim_{n \to \infty} n^{-1} \ln Z$.
For the thermodynamic limit of a graph which is a section of a regular lattice
graph $\Lambda$ with some boundary conditions, we denote $\kappa_\Lambda$ as
the coordination number satisfying
$\kappa_\Lambda = 2 \lim_{n(G) \to \infty} \ e(G)/n(G)$ 
in this limit (independent of boundary conditions).  Reviews of the Potts model
include \cite{wurev,baxterbook}. 

Let $G^\prime=(V,E^\prime)$ be a spanning subgraph of $G$, i.e. a subgraph
having the same vertex set $V$ and a subset of the edge set, $E^\prime
\subseteq E$. Then a useful representation of $Z(G,q,v)$ is \cite{kf}
\beq
Z(G,q,v) = \sum_{G^\prime \subseteq G} q^{k(G^\prime)}v^{e(G^\prime)}
\label{cluster}
\eeq
where $k(G^\prime)$ denotes the number of connected components of $G^\prime$.
Since we only consider connected lattice graphs $G$, we have $k(G)=1$.  For the
ferromagnet, eq. (\ref{cluster}) allows one to generalize $q$ from the positive
integers to the positive reals, ${\mathbb R}_+$.  As is evident from
eq. (\ref{cluster}), $Z(G,q,v)$ is a polynomial in $q$ with minimal and maximal
degrees 1 and $n(G)$, and a polynomial in $v$ with minimal and maximal degrees
0 and $e(G)$.  Thus, for a finite $G$, $Z(G,q,v)$ is completely determined by
its zeros and can be written as $Z(G,q,v) =\prod_{i=1}^{e(G)}[v-v_i(q)]$.

We first recall some previous related work. For the special case $q=2$ on the
square lattice, where one has the exact Onsager solution for the partition
function, it was found that (in the thermodynamic limit) the locus ${\cal B}$
consists of the union of circles $|a \pm 1|=\sqrt{2}$ \cite{fisher}.  The locus
${\cal B}$ has also been determined exactly for $q=2$ on various other $d=2$
lattices (e.g., \cite{abe,cmo}).  However, aside from the special case $q=2$,
$d=2$, one does not, in general, know ${\cal B}$ for lattice graphs of
dimensionality $d \ge 2$.  Indeed, studies of complex-temperature zeros of the
Potts model partition function for finite sections of various lattices have
found that they exhibit a considerable scatter, especially in the half-plane
with $Re(a) < 0$ \cite{mm}-\cite{kc}.  A complementary approach has been to
obtain exact solutions for the partition function and finite-width strips of
lattices with various boundary conditions and hence determine exactly the locus
${\cal B}$ in the infinite-length limit (e.g., \cite{a}-\cite{ts}.  These
systems are quasi-one-dimensional, so that ${\cal B}$ does not cross the
positive real $a$ axis, i.e., the free energy is analytic for all nonzero
temperatures.  With the definition $x_{sq} = v/\sqrt{q}$ for the square
lattice, it was found the complex-temperature zeros calculated for finite
sections of this square lattice with duality-preserving boundary conditions
have the property that a subset lies exactly on an arc of the unit circle
$|x_{sq}|=1$ in the complex $x_{sq}$ plane for $Re(x_{sq}) \gsim 0$
\cite{mbook,chw,wuetal,sdg}.  For physical temperature the coefficients of
powers of $a=e^K$ are positive, so there are no zeros on the positive real axis
$Re(a) > 0$ for any finite lattice. However, in the thermodynamic limit, the
phase boundary on ${\cal B}$ crosses this axis at the ferromagnetic phase
transition point, $a_c=1+\sqrt{q}$, i.e. $x_c=1$.  A number of interesting
results have been obtained by F. Y. Wu and collaborators: in Ref. \cite{chw} a
conjecture was made that for finite sections of the square lattice with
self-dual boundary conditions and for the same lattice with free or periodic
boundary conditions in the thermodynamic limit, the complex-temperature zeros
of the partition function in the $Re(x_{sq}) > 0$ half-plane are located on the
unit circle $|x_{sq}|=1$.  In Ref. \cite{wuetal}, it was shown, using Euler's
identity for integer partitions, that for the Potts model on the self-dual
square lattice, if one takes the limit $q \to \infty$, all of the zeros of the
partition function are located on this unit circle $|x_{sq}|=1$ for any lattice
size.  In Refs. \cite{wuetal,wu3,huangwu} it was shown that partition function
for the Potts model on a finite $d$-dimensional Cartesian lattice ${\mathbb
E}^d$ with $d \ge 2$ and appropriate boundary conditions, in the limit $q \to
\infty$, divided by an appropriate power of $q$, is given by the generating
function of restricted integer partitions in $d-1$ dimensions, with expansion
variable $t=v^d/q$.

In the present paper we present some results on the zeros of the Potts model
partition function for large $q$ on general regular $d$-dimensional lattices,
both in the thermodynamic limit and for finite lattices, with various boundary
conditions.  We shall give a simple proof for an arbitrary regular lattice of
dimension $d \ge 2$ that, when the thermodynamic limit and the $q \to \infty$
limit are taken in an appropriate way, the zeros of the partition function lie
on the circle $|x_\Lambda|=1$.  This generalizes the corresponding result of Wu
and coworkers from a Cartesian lattice ${\mathbb E}^d$ to an arbitrary regular
$d$-dimensional lattice. We also consider finite sections of some regular
lattices and determine the circular loci near which the zeros lie for large 
$q$.  We shall often call this the ``approximating circle'' for the zeros.

\section{General Results for Large $q$}
\label{sectionII}

In this section we analyze the pattern of partition function zeros of the Potts
model for large $q$.  Let us first arrange the terms in eq. (\ref{cluster}) in
order of decreasing power of $v$, from $v^{e(G)}$ to 1.  There are ${e(G)
\choose j}$ terms in the coefficients of $v^{e(G)-j}$ and $v^j$, so that in the
special case $q=1$ the partition function reduces to $Z(G,q=1,v)=(v+1)^{e(G)}$.
These ${e(G) \choose j}$ terms can have different powers in $q$ except for very
small or very large $j$, as follows.  The term with the highest power of $v$
arises from the contribution of $G'=G$ in eq. (\ref{cluster}), where all of the
edges of $G$ are present, and the power of $v$ is $e(G)$; since this graph
consists of a single connected component, the coefficient is $q$ and the term
is $qv^{e(G)}$.  In considering other terms in $Z(G,q,v)$, we assume that the
graph $G$ has the property that the removal of a few edges does not cut it into
disconnected components.  This excludes the cases of a one-dimensional lattice
and quasi-one-dimensional lattice strips of narrow widths; we comment on these
in the appendix.  For example, for a regular $d$-dimensional lattice graph, we
require that $d \ge 2$ and that the lengths $L_k$ with $1 \le k \le d$ are
great enough so that this condition is satisfied.  Next, the term in $Z(G,q,v)$
with one lower power of $v$ is obtained from the set of $G^\prime$ each of
which has an edge set $E'$ with one less edge than the full edge set $E$. This
term is ${e(G) \choose 1} \, q \, v^{e(G)-1}$, where ${\ell \choose
m}=\ell!/[m!(\ell-m)!]$.  Similarly, the first few terms with descending powers
of $v$ below the maximal power are given by ${e(G) \choose j} \, q \,
v^{e(G)-j}$, where $j$ edges are removed in $G^\prime$ and the linear
dependence on $q$ reflects the property that this edge removal maintains the
connectedness of the resulting $G'$, so $k(G')=1$. In general, this holds for
$j < \delta(G)$, where $\delta(G)$ and $\Delta(G)$ denote the minimal and
maximal degrees of a vertex in $G$. (The degree of a vertex is defined as the
number of edges connected to it).  Thus, for example, a term involving $q^2$
begins to appear in the coefficient of $v^{e(G)-3}$ for cyclic strips of the
square lattice with $L_1 \equiv L_x > 2$ and $L_2 \equiv L_y \ge 2$ since the
degree of vertices on the upper and lower boundaries is three, but it appears
in the coefficient of $v^{e(G)-2}$ for the same strip with $L_x=2$. At the
other end of the polynomial $Z(G,q,v)$ ordered as powers of $v$, the term with
the power of $v$ equal to zero arises from the contribution of $G'$ with
$E'=\emptyset$, i.e., the spanning subgraph containing no edges, leaving the
$n(G)$ vertices as disconnected components; this term is thus $q^{n(G)}$. The
term linear in $v$ results from the contributions of spanning subgraphs
$G^\prime$ with one edge and is $e(G) \, q^{n(G)-1} \, v$. In general, the
first few such terms are given by ${e(G) \choose j} \, q^{n(G)-j} \, v^j$,
where $G^\prime$ consist of $j$ edges such that $j < g(G)$, with $g(G)$
denoting the girth of $G$, i.e., the length of minimum closed circuit on $G$.
For example, for cyclic strips of the square lattice again, the term
$q^{n(G)-3}$ appears first in the coefficient of $v^4$ for $L_x > 3$ because
the girth is four, but term $q^{n(G)-2}$ appears first in the coefficient of
$v^3$ if $L_x=3$ and the term $q^{n(G)-1}$ appears in the coefficient of $v^2$
if $L_x=2$. Thus, for graphs $G$ satisfying the above-mentioned conditions, we
have
\beqs
& & Z(G,q,v) = \sum_{j=0}^{e(G)} c_j(q) v^j \cr\cr
& = & qv^{e(G)} + e(G)qv^{e(G)-1} + {e(G) \choose 2}qv^{e(G)-2} + ... \cr\cr
& + & {e(G) \choose 2}q^{n(G)-2}v^2 + e(G)q^{n(G)-1}v + q^{n(G)} \cr\cr
& & 
\label{Zexp}
\eeqs
where the coefficients $c_j(q)$ are polynomials in $q$.  

Now let us consider the limit $q \to \infty$.  Evidently, in this limit, for a
given finite graph $G$, there is a single dominant term in $Z(G,q,v)$, namely
$q^n$, arising from $G'$ with $E'=\emptyset$.  This term is what one would get
in the evaluation of $Z$ if there were no spin-spin interactions, but instead
just single-site contributions.  This reduction helps to understand the fact
that the Potts model is exactly solvable \cite{pg} in the limit $q \to \infty$.

Next, let us examine the terms in $Z(G,q,v)$ in more detail.  For a given graph
$G$, we define 
\beq
x_G = \frac{v}{q^{n(G)/e(G)}}
\label{xg}
\eeq
and, in the thermodynamic limit, 
\beq
x_\Lambda = \frac{v}{q^{2/\kappa_\Lambda}} \ . 
\label{x}
\eeq
Considering $q$ to be large, we focus on the first two and last two of the
terms in eq. (\ref{Zexp}); we can write these as 
\beqs
\lefteqn{Z(G,q,v)} \cr\cr
& = & q^{n(G)+1}\Bigl [x_G^{e(G)}\, \{1+e(G)[x_G \, q^{n(G)/e(G)}]^{-1} \} 
+ ... \cr\cr
& + & q^{-1} \, \{ e(G) \, x_G \, q^{(n(G)/e(G))-1} + 1 \} \Bigr ] \ . 
\label{Zexpx}
\eeqs
We can approximate $Z(G,q,v)$ by including just the first and last terms if
several conditions are met.  In the expression multiplying $x_G^{e(G)}$, the
second term is negligible compared with the first if $x_G \sim O(1)$ and the
condition
\beq
q \gg e(G)^{e(G)/(n(G)-1)}
\label{qcon1}
\eeq
is satisfied.  In the expression multiplying $q^{-1}$, the first term is
negligible compared with the second if $x_G \sim O(1)$ and the condition
\beq
q \gg [e(G)]^{e(G)/(e(G)-n(G)+1)}
\label{qcon2}
\eeq
is satisfied. Let us denote the degree of $c_j(q)$ in $q$ as $p_j$. Among the
${e(G) \choose j}$ terms in $c_j(q)$, only a few have this power except
for small and large $j$ as discussed above. Now we require that 
$c_j(q)v^j$ be small compared with $q^n$, so that in terms of the
power of $q$,
\beq
p_j + j \left ( \frac{n(G)-1}{e(G)} \right ) < n(G)
\label{pj1}
\eeq
or
\beq
p_j < \frac{(e(G)-j)(n(G)-1)}{e(G)} + 1 \ .
\label{pj2} 
\eeq
Further, if these conditions hold and $x_G \sim O(1)$, then terms arising from
the third, fourth, and subsequent terms of $Z(G,q,v)$ are small relative to the
first, and the third, fourth, and earlier terms counting in from the last, are
negligible relative to the last, with the ordering specified by ascending
powers of $v$, as in eq. (\ref{Zexp}).  Provided that $x_G \sim O(1)$ and the
above conditions hold, one can approximate $Z(G,q,v)$ by keeping only the first
and last terms in eq. (\ref{Zexp}):
\beq
Z(G,q,v) \sim qv^{e(G)} + q^{n(G)} = q^{n(G)+1}[x_G^{e(G)}+q^{-1}] \ .
\label{Zapprox}
\eeq
The zeros of this approximation to $Z(G,q,v)$ are given by
\beq 
x_G = (-q^{-1})^{1/e(G)} \ . 
\label{xgeq}
\eeq
Now consider the limit $e(G) \to \infty$ and $n(G) \to
\infty$ with the ratio $e(G)/n(G)=\kappa_G/2$ finite.  If $q$ also goes to
infinity, sufficiently fast to satisfy the conditions (\ref{qcon1}) and
(\ref{qcon2}) but more slowly than the exponential $e^{b e(G)}$ with
$b$ a real positive constant, then 
\beq
\lim_{q \to \infty} \lim_{e(G) \to \infty} \,  |q|^{1/e(G)} = 1 \ ,
\label{qecon}
\eeq
and these zeros merge onto the unit circle
\beq
|x_G|=1 \ .
\label{xg1}
\eeq
If $q$ were to grow more rapidly, as $q \sim e^{b e(G)}$ where $b$ is an
arbitrary positive real constant, then the zeros would merge onto the circle
$|x_G|=e^{-b}$.

In particular, one may consider the thermodynamic limit of a graph $G$ which is
a section of the regular lattice $\Lambda$.  Then the conditions 
corresponding to eqs. (\ref{qcon1}) and (\ref{qcon2}) are, respectively,
\beq
q \gg e(G)^{\kappa_\Lambda/2} 
\label{qcon1p}
\eeq
and
\beq
q \gg e(G)^{\kappa_\Lambda/(\kappa_\Lambda-2)} \ . 
\label{qcon2p}
\eeq
The inequality $\kappa_\Lambda/2 \ge \kappa_\Lambda/(\kappa_\Lambda-2)$ is
equivalent to the inequality $\kappa_\Lambda \ge 4$.  Since $\kappa_\Lambda \ge
4$ for all of the lattices that we consider except for the honeycomb lattice,
a consequence is that for these former lattices with $\kappa_\Lambda \ge 4$,
condition (\ref{qcon1p}) implies condition (\ref{qcon2p}).  Provided that
$x_\Lambda \sim O(1)$ and the above conditions hold, so that one can
approximate $Z(G,q,v)$ with the first and last terms in eq. (\ref{Zexp}), it
follows that in this thermodynamic limit, with $q$ also going to infinity
sufficiently fast to satisfy the conditions (\ref{qcon1p}) and (\ref{qcon2p})
but more slowly than the exponential form given above, the zeros of $Z(G,q,v)$
merge onto the circle
\beq
|x_\Lambda|=1
\label{xcircle}
\eeq
where $x_\Lambda$ is given by eq. (\ref{x}).  Our result (\ref{xcircle}) holds
for any regular lattice with dimension $d \ge 2$.  For example, for the
$d$-dimensional Cartestian lattice ${\mathbb E}^d$, $\kappa_{{\mathbb
E}^d}=2d$, while for the $d$-dimensional body-centered cubic lattice,
$\kappa_{bcc^d}=2^d$.

Since the free energy is nonanalytic at $x_\Lambda=1$, eq. 
(\ref{xcircle}) yields the asymptotic relations
\beq
v_{c,\Lambda} \sim q^{2/\kappa_\Lambda} \quad {\rm as} \ \ q \to \infty
\label{vc}
\eeq
and hence 
\beq
K_{c,\Lambda} \sim \frac{2}{\kappa_\Lambda} \ln q  \quad {\rm as} \ \ 
q \to \infty
\label{kc}
\eeq
for the values of $v$ and $K$ where the Potts model on the lattice $\Lambda$
has a phase transition from a paramagnetic (PM) high-temperature phase to a
ferromagnetic (FM) low-temperature phase in the limit of large
$q$.  We note that this agrees with the large-$q$ limit of the mean-field
theory result \cite{mft}
\beq
K_{c,MFT} = \frac{2(q-1)}{\kappa_\Lambda(q-2)} \ \ln (q-1) \ , 
\label{kc_mft}
\eeq
viz., 
\beqs
K_{c,MFT} & \sim & \frac{2}{\kappa_\Lambda} \Bigl [ 
\{ 1 + q^{-1} + O(q^{-2}) \} \ln q \cr\cr
  & - & \{ q^{-1} + O(q^{-2}) \} \Bigr ] 
\label{kc_mftseries}
\eeqs
as $q \to \infty$.  In eqs. (\ref{kc}) and (\ref{kc_mft}), the property that
$K_c$ increases with increasing $q$ can be understood as a consequence of the
fact that as $q$ gets large, each spin has more possible values (is
``floppier''), and hence one must go to a lower temperature for the
ferromagnetic long-range order to occur.  The feature that $K_c$ increases
asymptotically like $\ln q$ as $q \to \infty$ can be understood since an
order-disorder transition involves a balance between minimizing the
configurational energy and maximizing the entropy terms in the free energy per
site, $F=U-TS$, and in this limit, the entropy per site is $S \to k_B \ln q$.

Next, we recall the known exact equations for the PM-FM phase transition on the
2D lattices \cite{wurev,baxterbook}, which we write in a convenient manner for
the discussion of the large-$q$ limit, 
\beqs
\frac{q}{v^2}=1 & & \qquad {\rm for} \ \ \Lambda=sq \cr\cr
\frac{q}{v^3}=1+\frac{3}{v} & & \qquad {\rm for} \ \ \Lambda=tri \cr\cr
\frac{q^2}{v^3}=1-\frac{3q}{v^2} & & \qquad {\rm for} \ \ \Lambda=hc \ ,
\eeqs
where $sq$, $tri$, and $hc$ denote the square, triangular, and honeycomb
lattices.  As is evident, in the limit of large $q$ and $v$ 
such that $q=v^2$ for the square lattice, $q=v^3$ for the triangular
lattice and $q=v^{3/2}$ for the honeycomb lattice, these equations are all in
agreement with the general relation (\ref{vc}), which also holds for
higher-dimensional lattices.

\section{Partition Function Zeros for Finite Lattice Sections} 
\label{sectionIII}

\subsection{General Structure} 

It is also of interest to consider zeros of $Z(G,q,v)$ on finite lattice graphs
in the limit of large $q$.  In general, if the conditions specified in the
previous section are satisfied, one can approximate $Z(G,q,v)$ by keeping the
first and last terms in eq. (\ref{Zexp}).  We have studied these zeros in
detail for sections of various 2D lattices and have found that a sufficient
criterion that these conditions can be satisfied for these and
higher-dimensional lattices is that one uses fully periodic boundary
conditions.  Accordingly, when $q$ is large, the zeros in the $x_{\Lambda}$
plane are located close to a circle.  For finite lattice graphs with fully
periodic boundary conditions, $e(G)/n(G)=\kappa_\Lambda/2$ so
that the two definitions in eqs.(\ref{xg}) and (\ref{x}) are
equivalent. However, for finite graphs with other boundary conditions, $x_G$
varies from one graph to another. Therefore, we will plot zeros in the
$x_\Lambda$ plane for the lattice $\Lambda$ with different boundary
conditions. We first rewrite the approximate expression (\ref{Zapprox}) in an
equivalent form that will be convenient for our discussion; when $q$ is large,
\beqs
& & Z(G,q,v) \sim qv^{e(G)} + q^{n(G)} \cr\cr
& = & q^{(2e(G)/\kappa_\Lambda)+1}
[x_\Lambda^{e(G)}+q^{n(G)-(2e(G)/\kappa_\Lambda)-1}]
\label{Zapprox2}
\eeqs
The zeros of this approximation to $Z(G,q,v)$ are located
on a circle in the $x_\Lambda$ plane with radius 
\beq
r(\Lambda,BC) = q^{p(\Lambda,BC)}
\label{rcircle}
\eeq
where
\beq
p(\Lambda,BC)=\frac{n(G)-1}{e(G)}-\frac{2}{\kappa_\Lambda} \ .
\label{ppower}
\eeq
(This power $p$ should not be confused with the powers $p_j$ in
eqs. (\ref{pj1}), (\ref{pj2}).  The radius $r(\Lambda,BC)$ is less (greater)
than unity if $p(\Lambda,BC)$ is negative (positive). For lattices with
periodic boundary conditions in all directions, $n(G)/e(G)=2/\kappa_\Lambda$
and hence
\beq
p(\Lambda,PBC)=-\frac{1}{e(G)}
\label{ppbc}
\eeq
as in eq. (\ref{xgeq}).

In our explicit calculations of zeros for various lattices, we have found that
for large but finite $q$, the approximating circles near which these zeros lie
may have centers that are slightly shifted to the left or right of the origin 
of the $x_\Lambda$ plane and move in toward the origin as $q \to \infty$. 
The following analysis provides an understanding of these shifts.  As
discussed in Section \ref{sectionII}, there are $\delta(G)$ terms in the
partition function given by ${e(G) \choose j}qv^{e(G)-j}$ and there are $g(G)$
terms given by ${e(G) \choose j}q^{n(G)-j}v^j$. For the finite lattice graphs
with large vertex degree and small girth, a specific approximation for the
partition function $Z(G,q,v)$ when $q$ is large is
\beq
Z(G,q,v) \sim q(v+1)^{e(G)}+q^{n(G)}
\label{zlargedegree}
\eeq
so that the center of the circle, denoted by $c(\Lambda,BC)$, is
not at the origin but at
\beq
c(\Lambda,BC)=-\frac{1}{q^{2/\kappa_\Lambda}} \ .
\label{c_left} 
\eeq
Of course, this approaches zero as $q \to \infty$, in agreement with our
result above, eq. (\ref{xcircle}).  The triangular lattice is an example 
for which the center of the circle is shifted to the left, as in
eq. (\ref{c_left}) and will be discussed below. 

On the other hand, for finite lattice graphs with small vertex degree and 
large girth, a specific approximation for $Z(G,q,v)$ when $q$ is large is 
\beqs
& & Z(G,q,v) \sim qv^{e(G)}+q^{n(G)-e(G)}(q+v)^{e(G)} \cr\cr
& = & q^{n(G)+1}[x_G^{e(G)}+q^{-1}(1+q^{(n(G)/e(G))-1}x_G)^{e(G)}] \cr\cr
& & 
\label{zlargegirth} 
\eeqs
The circle near which the zeros lie is now shifted to the right. Let
us first consider the case with fully periodic boundary conditions so that
$x_G = x_\Lambda$. In the complex $x_\Lambda$
plane, the circle has radius 
\beq
r(\Lambda,PBC)=\frac{q^{1/e(G)}}{q^{2/e(G)}-q^{(2n(G)/e(G))-2}}
\label{rlargegirth}
\eeq
and center at
\beq
c(\Lambda,PBC)=\frac{q^{n/e(G)-1}}{q^{2/e(G)}-q^{(2n(G)/e(G))-2}} \ .
\label{c_right}
\eeq
For the lattices with $e(G)$ larger than $n(G)$, the term $q^{(2n(G)/e(G))-2}$ in the
denominator is negligible when $q$ is large, and the radius of the circle is
approximately $q^{-1/e(G)}$, as in eq. (\ref{rcircle}). For large $q$, the 
center of the circle in the $x_\Lambda$ plane is 
\beq
c(\Lambda,PBC)=q^{-s(\Lambda,BC)}
\eeq
where 
\beq
s(\Lambda,PBC)=1+\frac{2-n(G)}{e(G)} \ .
\label{cqpowerpbc}
\eeq
When $e(G)$ is large, so that $2/e(G) \ll 1$, the position of center can be
further approximated as $c(\Lambda,PBC) \simeq q^{(2/\kappa_\Lambda)-1}$,
which approaches the origin as $q \to \infty$.

For the lattice sections with other boundary conditions, a specific
approximation for $Z(G,q,v)$ for large $q$ is 
\beqs
& & Z(G,q,v) \sim q^{(2e(G)/\kappa_\Lambda)+1} \times \cr\cr
& & [x_\Lambda^{e(G)}+
q^{n(G)-(2e(G)/\kappa_\Lambda)-1}(1+q^{(2/\kappa_\Lambda)-1}x_\Lambda)^{e(G)}]
\cr\cr
& & 
\eeqs
In the $x_\Lambda$ plane, the resultant approximating circle has
radius 
\beq
r(\Lambda,BC)=\frac{q^{1/e(G)-(n(G)/e(G))+2/\kappa_\Lambda}}
{q^{2/e(G)-(2n(G)/e(G))+(4/\kappa_\Lambda)}-q^{(4/\kappa_\Lambda)-2}}
\eeq
and center 
\beq
c(\Lambda,BC)=\frac{q^{(2/\kappa_\Lambda)-1}}
{q^{2/e(G)-(2n(G)/e(G))+(4/\kappa_\Lambda)}-q^{(4/\kappa_\Lambda)-2}}
\ .
\eeq
For lattices with $\kappa_\Lambda>2$, $q^{4/\kappa_\Lambda-2}$ in the
denominator is negligible when $q$ is large, and the radius of the circle is
the same as eqs. (\ref{rcircle}) and (\ref{ppower}). We also have
\beq
s(\Lambda,BC)=1+\frac{2-2n(G)}{e(G)}+\frac{2}{\kappa_\Lambda}
\label{cqpower}
\eeq
for large $q$.  The honeycomb lattice is an example for which the circle near
which the zeros lie shifts to the right of the origin and will be discussed
below. Another example is the square lattice with next-nearest-neighbor
interactions that have the same strength as the nearest-neighbor interactions,
which we have also analyzed. From the above discussion, we expect that the
center of the circle is the origin for lattice sections with vertex degree
roughly equal to girth.

\subsection{Square Lattice Sections}

\begin{figure}[hbtp]
\epsfxsize=1.5in
\epsffile{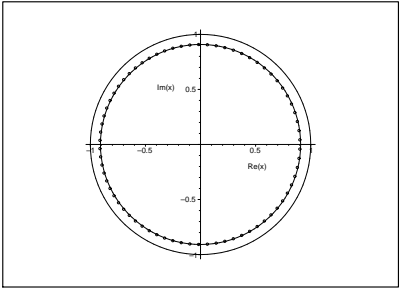}
(a) \\
\epsfxsize=1.5in
\epsffile{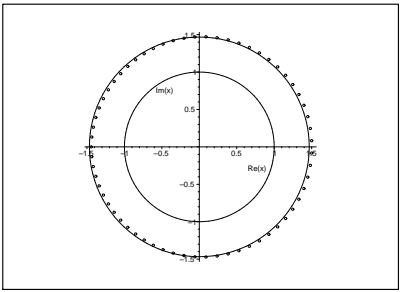}
(b) \\
\epsfxsize=1.5in
\epsffile{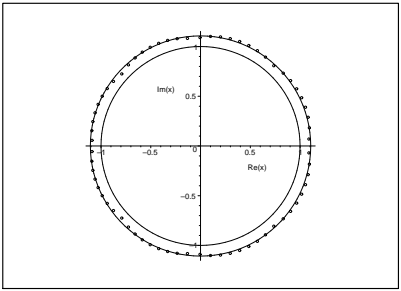}
(c) \\
\epsfxsize=1.5in
\epsffile{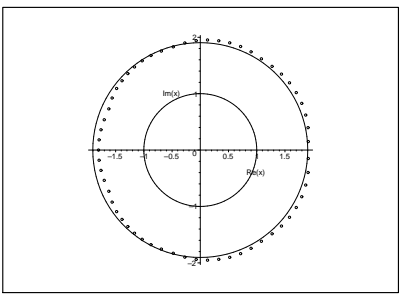}
(d) \\
\caption{\footnotesize{Zeros of the Potts model partition function, plotted in
the $x_{sq}$ plane, for a section of the square lattice with $L_y=4$ and
$L_x=9$, when $q=1000$, for the following boundary conditions: (a) toroidal;
(b) cyclic; (c) cylindrical; (d) free. For comparison, the unit circle and
circles where the zeros cluster are also shown.}}
\label{sqq1000}
\end{figure}

For the square lattice with toroidal (tor) boundary conditions,
$e(sq,tor)=2L_xL_y$, $\kappa_{sq} = 4$, and
\beq
p(sq,tor)=-\frac{1}{2L_xL_y} \ .
\label{psqtor}
\eeq
Hence the radius of the approximating circle increases as the area of the
lattice increases, and approaches unity from below in the thermodynamic
limit. The zeros of the partition function for the Potts model on the toroidal
strip of the square lattice with $L_y=4$ and $L_x=9$ in the $x_{sq}$ plane when
$q=1000$ are shown in Fig. \ref{sqq1000}(a). The radius of the circle is about
0.9085. The zeros lie very close to this circle for the above value of $q$, and
we find that they get closer to the circle given by eqs. (\ref{rcircle}) and
(\ref{ppower}) when $q$ increases.

For the square lattice with cyclic boundary conditions we have
$e(sq,cyc)=L_x(2L_y-1)$. By eq.(\ref{ppower}), 
\beq
p(sq,cyc)=\frac{L_x-2}{2L_x(2L_y-1)} \ . 
\eeq
In the thermodynamic limit the radius of the approximating circle decreases and
approaches unity from above. Note that if one were to keep $L_y$ fixed and
take $L_x \to \infty$, thereby violating the premises of our analysis, the
radius of this approximating circle would be $q^{1/2(2L_y-1)}$, which would
diverge as $q \to \infty$. The zeros of the cyclic strip with $L_y=4$ and
$L_x=9$ in the $x_{sq}$ plane when $q=1000$ are shown in
Fig. \ref{sqq1000}(b). The radius of the approximating circle is about
1.47. The zeros are not as close to the circle as for the corresponding
toroidal case. On the other hand, if one keeps $L_x$ fixed and takes $L_y \to
\infty$, corresponding to cylindrical boundary conditions, the radius of the
approximating circle approaches unity from above. The zeros of the cylindrical
strip with $L_y=4$ and $L_x=9$ in the $x_{sq}$ plane when $q=1000$ are shown in
Fig. \ref{sqq1000}(c). The radius of this approximating circle is about
1.11. The zeros are closer to the circle than the cyclic strip with the same
$L_y$ and $L_x$ because the number of boundary vertices (i.e. vertices with
coordination number three) is reduced.

For the square lattice with free boundary conditions, 
$e(sq,free)=2L_xL_y-L_x-L_y$. By eq.(\ref{ppower}), 
\beq
p(sq,free)=\frac{L_x+L_y-2}{2(2L_xL_y-L_x-L_y)} \ . 
\eeq
In the thermodynamic limit the radius of the circle decreases and approaches
unity from above. Again, however, if one were to keep $L_y$ fixed and take $L_x
\to \infty$, the radius of the approximating circle would be $q^{1/2(2L_y-1)}$,
(the same as the cyclic case), which would diverge as $q \to \infty$. Since for
a given lattice size the number of boundary vertices is larger than for other
boundary conditions, and especially since there are four vertices with
coordination number equal to two, it is natural to expect that the zeros will
lie farther from the asymptotic circle than for these other boundary
conditions. The zeros of the free strip with $L_y=4$ and $L_x=9$ in the
$x_{sq}$ plane when $q=1000$ are shown in Fig. \ref{sqq1000}(d). The radius of
the approximating circle is about 1.90.

\subsection{Triangular lattice}

\begin{figure}[hbtp]
\epsfxsize=1.5in
\epsffile{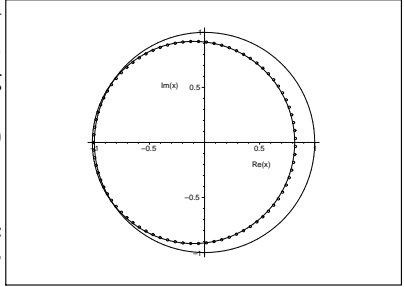}
(a) \\
\epsfxsize=1.5in
\epsffile{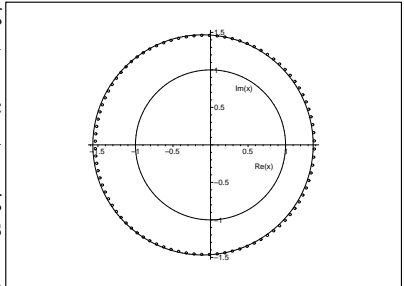}
(b) \\
\epsfxsize=1.5in
\epsffile{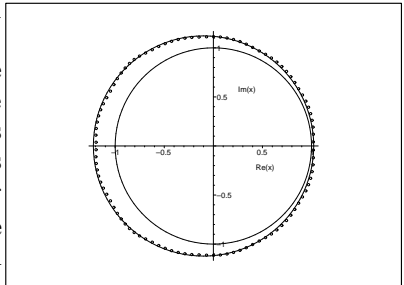}
(c) \\
\epsfxsize=1.5in
\epsffile{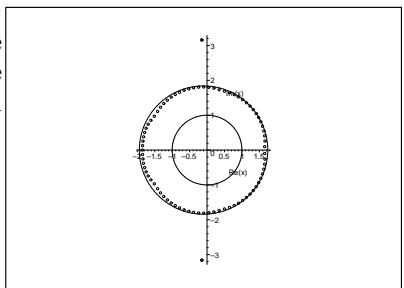}
(d) \\
\caption{\footnotesize{Zeros of the Potts model partition function in the
$x_{tri}$ plane for the triangular lattice when $q=1000$, for the following
boundary conditions and sizes: (a) toroidal with $L_y=3$ and $L_x=9$; (b)
cyclic with $L_y=4$ and $L_x=9$; (c) cylindrical with $L_y=4$ and $L_x=9$; (d)
free with $L_y=4$ and $L_x=9$. For comparison, the unit circle and
approximating circles are also shown.}}
\label{triq1000}
\end{figure}

For the triangular lattice with toroidal boundary conditions,
$e(tri,tor)=3L_xL_y$ and $\kappa(tri,tor)=6$.  Hence, for a finite section of
the triangular lattice with toroidal boundary conditions, as for the infinite
triangular lattice,
\beq
x_{tri}=\frac{v}{q^{1/3}}
\eeq
and 
\beq
p(tri,tor)=-\frac{1}{3L_xL_y} \ , 
\eeq
as given by eq.(\ref{xgeq}). This radius increases as the size of the lattice
increases, and approaches unity from below in the thermodynamic limit. There is
a noticeable shift of the circle to the left. The center of the approximating
circle is at $x_{tri}=-q^{-1/3}$, approaching zero as $q \to
\infty$. The zeros of the toroidal strip with $L_y=3$ and $L_x=9$ in the
$x_{tri}$ plane when $q=1000$ are shown in Fig. \ref{triq1000}(a). The radius
of this approximating circle is about 0.918.

For a section of the triangular lattice with cyclic boundary conditions, 
$e(tri,cyc)=L_x(3L_y-2)$. When $q$ is large, we calculate
\beq
p(tri,cyc)=\frac{2L_x-3}{3L_x(3L_y-2)}
\eeq
for the approximating circle. In the thermodynamic limit, the radius of this
circle decreases, approaching unity from above. However, if one were to keep
$L_y$ fixed and take $L_x \to \infty$, then the radius would be 
$q^{2/3(3L_y-2)}$, which would diverge as $q \to \infty$. The zeros of the
cyclic strip with $L_y=4$ and $L_x=9$ in the $x_{tri}$ plane when $q=1000$ are
shown in Fig. \ref{triq1000}(b). The radius of this approximating circle is
about 1.47 as for the corresponding case of the square lattice. 
If one keeps $L_x$ fixed and takes $L_y \to \infty$, corresponding to
cylindrical boundary conditions, the radius of the approximating circle 
approaches unity from above. The zeros of the cylindrical strip with $L_y=4$
and $L_x=9$ in the $x_{tri}$ plane when $q=1000$ are shown in
Fig. \ref{triq1000}(c). The radius of this approximating circle is about 1.12.

In general, the approximation in eq.(\ref{Zapprox}) is not valid for the
triangular lattice with free boundary conditions. The reason is that there are
two vertices with degree two, so that the coefficient of $v^{e(tri,free)-2}$
contains the term $2q^2$ and the coefficient of $v^{e(tri,free)-4}$ contains
the term $q^3$ for $L_x, L_y > 2$. Therefore, when $q$ is large there are two
pair of roots around $v \sim \pm i \sqrt{q}$, or equivalently around $\pm
i q^{1/6}$ in the $x_{tri}$ plane.  We have $e(tri,free)=3L_xL_y-2L_x-2L_y+1$.
The partition function zeros for the free strip of the triangular lattice with
$L_y=4$ and $L_x=9$ when $q=1000$ are shown in
Fig. \ref{triq1000}(d).  Although the approximation (\ref{Zapprox}) does not
apply here, one can see that many of these zeros lie close to an 
approximating circle with radius 1.84, which is similar to what one would get
by formally using $p(tri,free)=2(L_x+L_y-2)/[3(3L_xL_y-2L_x-2L_y+1)]$. 
We observe that there are two almost overlapping complex-conjugate pairs
of zeros which have relatively large magnitude and are close to the imaginary
axis.

\subsection{Honeycomb lattice}

\begin{figure}[hbtp]
\epsfxsize=1.5in
\epsffile{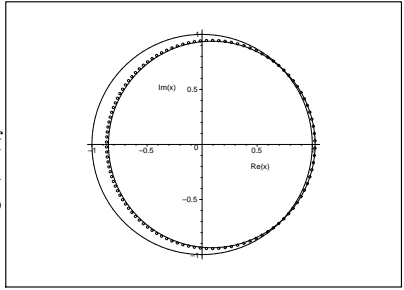}
(a) \\
\epsfxsize=1.5in
\epsffile{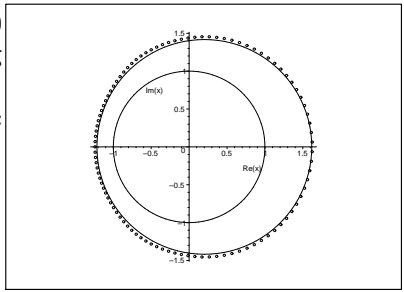}
(b) \\
\epsfxsize=1.5in
\epsffile{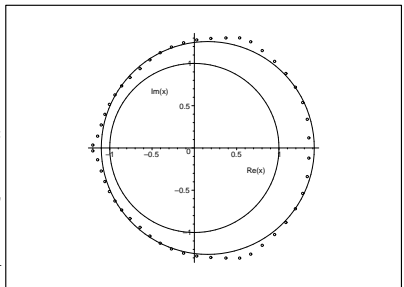}
(c) \\
\epsfxsize=1.5in
\epsffile{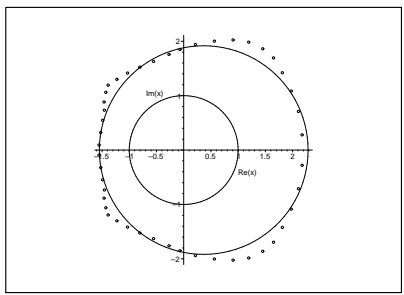}
(d) \\
\caption{\footnotesize{Zeros of the Potts model partition function in the
$x_{hc}$ plane for the honeycomb lattice when $q=1000$, for the following
boundary conditions and sizes: (a) toroidal with $L_y=4$ and $L_x=18$; (b)
cyclic with $L_y=4$ and $L_x=18$; (c) cylindrical with $L_y=4$ and $L_x=9$; (d) free boundary conditions with $L_y=4$ and $L_x=9$.
For comparison, the unit circle and approximating circles are also shown.}}
\label{hcq1000}
\end{figure}

For the honeycomb lattice with toroidal boundary conditions where both $L_x$
and $L_y$ are even, $e(hc,tor)=3L_xL_y/2$ so that $n(hc,tor)/e(hc,tor)=2/3$ for
any $L_x$ and $L_y$. Thus, 
\beq
x_{hc}=\frac{v}{q^{2/3}}
\eeq
and
\beq
p(hc,tor)=-\frac{2}{3L_xL_y}  \ , 
\eeq
as given in eq.(\ref{xgeq}). This radius increases as the size of the lattice
increases, and approaches unity from below in the thermodynamic limit. There is
a shift of the circle to the right. By eq.(\ref{cqpowerpbc}), 
\beq
s(hc,tor)=\frac{1}{3} + \frac{4}{3L_xL_y}
\eeq
which approaches $1/3$ as the size of the lattice section becomes large.  The
zeros of the toroidal strip with $L_y=4$ and $m=9 (L_x=18)$ in the $x_{hc}$
plane when $q=1000$ are shown in Fig. \ref{hcq1000}(a). The radius of this
approximating circle is about 0.938.

For a section of the honeycomb lattice with cyclic boundry conditions we have
$e(hc,cyc)=L_x(3L_y-1)/2$, where $L_x$ must be even. By eq.(\ref{ppower}), 
\beq
p(hc,cyc)=\frac{2(L_x-3)}{3L_x(3L_y-1)} \ . 
\eeq
In the thermodynamic limit, the radius of the approximating circle 
decreases, approaching unity from above. For the shift of the circle, by 
eq. (\ref{cqpower}), 
\beq
s(hc,cyc)=\frac{1}{3} + \frac{4(3-L_x)}{3L_x(3L_y-1)}
\eeq
The zeros of the cyclic strip with $L_y=4$ and $m=9 (L_x=18)$ in the $x_{hc}$
plane when $q=1000$ are shown in Fig. \ref{hcq1000}(b). The radius of this
approximating circle is about 1.42.

Since we represent the honeycomb lattice in the form of bricks oriented
horizontally, the results for cylindrical boundary conditions cannot be
obtained from cyclic boundary conditions by simply switching $L_x$ and
$L_y$. We have $e(hc,cyl)=L_y(3L_x/2-1)$ for the honeycomb lattice with
cylindrical boundary conditions where $L_y$ must be even. When $q$ is large, we
calculate
\beq
p(hc,cyl)=\frac{2(2L_y-3)}{3L_y(3L_x-2)} \ . 
\eeq
and 
\beq
s(hc,cyl)=\frac{1}{3} + \frac{4(3-2L_y)}{L_y(3L_x-2)} \ .
\eeq
The zeros of the cylindrical strip with $L_y=4$ and $L_x=9$ in the $x_{hc}$
plane when $q=1000$ are shown in Fig. \ref{hcq1000}(c). The radius of this
approximating circle is about 1.26.  We have also considered other boundary 
conditions, as shown in Fig. \ref{hcq1000}.  

\subsection{Kagom\'e lattice}

\begin{figure}[hbtp]
\epsfxsize=1.5in
\epsffile{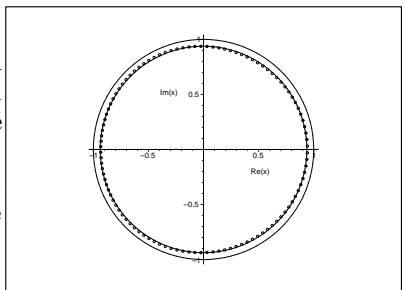}
(a) \\
\epsfxsize=1.5in
\epsffile{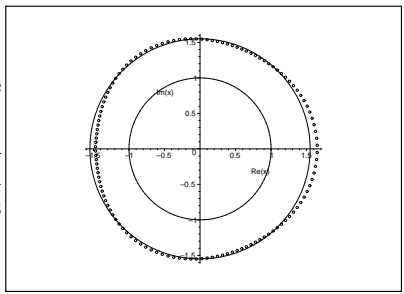}
(b) \\
\epsfxsize=1.5in
\epsffile{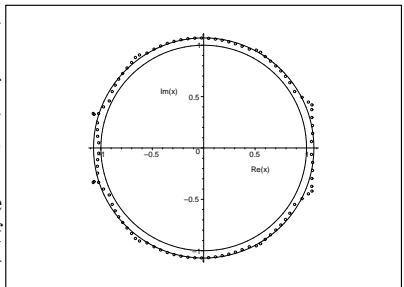}
(c) \\
\epsfxsize=1.5in
\epsffile{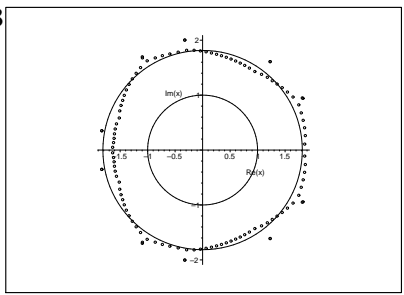}
(d) \\
\caption{\footnotesize{Zeros in the $x_{kag}$ plane for Potts model partition
function on the Kagom\'e lattice when $q=1000$, for the following boundary
conditions and sizes: (a) toroidal with $L_y=2$ and $L_x=9$; (b) cyclic with
$L_y=3$ and $L_x=9$; (c) cylindrical with $L_y=2$ and $L_x=9$; (d) free with
$L_y=3$ and $L_x=9$. For comparison, the unit circle and approximating circles
are also shown.}}
\label{kagq1000}
\end{figure}

As an example of a heteropolygonal two-dimensional lattice (an Archimedean
lattice comprised of more than one type of regular polygon), we consider the
kagom\'e lattice.  For a section of this lattice with toroidal boundary
conditions, $n(kag,tor)=3L_xL_y$ and $e(kag,tor)=6L_xL_y$ so that 
$n(kag,tor)/e(kag,tor)=1/2$ as for the square lattice.  Hence, 
\beq
x_{kag}=\frac{v}{\sqrt{q}}
\eeq
and
\beq
p(kag,tor)=-\frac{1}{6L_xL_y} \ . 
\eeq
The radius of the approximating circle increases as the size of the lattice
increases and approaches unity from below in the thermodynamic limit. The
zeros of the toroidal strip with $L_y=2$ and $L_x=9$ in the $x_{kag}$ plane
when $q=1000$ are shown in Fig. \ref{kagq1000}(a). The radius of this
approximating circle is about 0.938.  We have also considered other boundary
conditions as shown in Fig. \ref{kagq1000}.

\section{Summary}

In summary, we have presented some new results on the distribution of
complex-temperature zeros of the partition function of the $q$-state Potts
model for $q \to \infty$.  Generalizing previous work of Wu and coworkers for
Cartesian lattices, we have shown that with an appropriate definition of the
thermodynamic limit and the limit $q \to \infty$ on an arbitrary regular
lattice $\Lambda$ of dimensionality $d \ge 2$, the zeros lie on the circle
$|x_\Lambda|=1$.  We have also studied the distribution of zeros for finite
sections of various two-dimensional lattices for large $q$, showing how they
also lie approximately on circles.  

\bigskip

Acknowledgments: The research of R.S. was partially
supported by the NSF grant PHY-00-98527.  The research of S.C.C. was partially
supported by the NSC grant NSC-93-2119-M-006-009, NSC-94-2112-M-006-013 and NSC-94-2119-M-002-001. 

\newpage

\section{Appendix I: 1D Lattice and Quasi-1D Lattice Strips}

Our derivation of eq. (\ref{xcircle}) assumed that the lattice graph $G$ has
the property that removing a few edges does not cut it into disconnected parts.
This property does not hold for a 1D lattice or quasi-1D lattice strips of
sufficiently small width.  Here we comment briefly on this. We consider first a
one-dimensional lattice (line graph) with free or periodic boundary conditions,
labelled FBC and PBC, respectively.  We denote such graphs with $n$ vertices as
$L_{n,F}$ and $L_{n,P}$.  The coordination number $\kappa=2$ for $L_{n,P}$, and
this is the effective value of the coordination number also for $L_{n,F}$ when
$n \to \infty$, so that in both cases, $x_{1D}=v/q$.  From
elementary calculations one has the exact results
\beq
Z(L_{n,F},q,v) = q^n(1+x_{1D})^{n-1}
\label{ZL1f}
\eeq
and
\beq
Z(L_{n,P},q,v) = q^n[(1+x_{1D})^n + (q-1)x_{1D}^n]
\label{ZL1p}
\eeq
As in the text, we focus on large $n$ and large $q$.  Aside from the 
zero at $q=0$, which is not relevant in this case, the zeros of
$Z(L_{n,F},q,v)$ occur at the single point
\beq
x_{1D}=-1 \quad (FBC) \ .
\label{zeros_1d_fbc}
\eeq
Note that the number of edges for this graph is $e(L_{n,F})=n-1$.  Because the
degree in $q$ of the coefficient $c_j(q)$ defined in eq. (\ref{Zexp}) is
$p_j=n-j$, all terms have the same order and none can be neglected.

In the limit $n \to \infty$, followed by $q \to \infty$, the zeros of
$Z(L_{n,P},q,v)$ are given by the solution to the equation
$|1+x_{1D}|=|x_{1D}|$, which is the vertical line $Re(x_{1D})=-1/2$ in the
$x_{1D}$ plane, i.e.,
\beq
x_{1D} = -\frac{1}{2} + iy \quad (PBC) \ .
\label{zeros_1d_pbc}
\eeq
where $-\infty \le y \le \infty$. As these exact results show, the 1D case is
different from the thermodynamic limit of lattices with dimensionality $d \ge
2$ in that the locus of zeros is strongly dependent upon the choice of boundary
conditions and, moreover, is not the circle $|x_\Lambda|=1$.

In contrast, for strips of the square lattice with various types of boundary
conditions that maintain the duality of the infinite square lattice, we have
shown via exact solutions in Ref. \cite{sdg} that as $q \to \infty$, the zeros
do lie exactly on the unit circle $|x_\Lambda|=1$.  For these strips we also
found that for large but finite $q$, the locus ${\cal B}$ consists of the union
of (i) a self-conjugate portion of this unit circle given by
$x_{sq}=e^{i\theta}$ with $\theta_0 \le \theta \le \pi$ and $-\pi \le \theta
\le -\theta_0$ (where the angle $\theta_0$ depends on the strip) with (ii) a
line segment $-\rho \le x_{sq} \le -1/\rho$ on the negative real axis, where
$\rho > 1$ is a positive real constant (see the plots given in
Ref. \cite{sdg}).  The arc of the unit circle and the line segment intersect at
$x_{sq}=-1$.  As $q \to \infty$, $\rho \to 1$ and $\theta_0 \to 0$.  Thus, for
large but finite $q$, ${\cal B}$ is almost the unit circle except for a small
line segment centered at $x_{sq}=-1$ and a gap in the circle in the vicinity of
$x_{sq}=1$.  As we discussed in Ref. \cite{sdg}, these deviations can be
understood from the fact that the large-$q$ expansion of the dominant
eigenvalues of the transfer matrix contain poles at $x_{sq}=\pm 1$, showing
that regardless of how large $q$ is, the large-$q$ expansion breaks down at
these two points.

More generally, we have carried out similar large-$q$ expansions for the
dominant eigenvalues of the transfer matrix for quasi-one-dimensional strips of
several lattices with different boundary conditions.  We find that for the
strips we have considered, these expansions exhibit singularities on the unit
circle $x_\Lambda=1$ for strips with periodic transverse boundary conditions
but not for strips with free transverse boundary conditions.  For example, we
find that this expansion for the $L_y=2$ cylindrical strip of the square
lattice has branch-point singularities at $x_{sq}=\pm e^{i\pi/6}$ and
$x_{sq}=\pm e^{-i\pi/6}$.  Similarly, for the $L_y=2$ cylindrical strip of the
triangular lattice we find that the large-$q$ expansion of the dominant
eigenvalues has branch-point singularities at $x_{tri}=\pm 1$, $x_{tri}=\pm
e^{i \pi/3}$, and $x_{tri}= \pm e^{-i\pi/3}$.  

These deviations are characteristic of certain quasi-one-dimensional strips,
which do not satisfy the premises of our general analytic results in Section
II.  We do not find such deviations for the thermodynamic limits of sections of
regular lattices of dimensionality $d \ge 2$ in the limit of large $q$.

\vfill
\eject
\end{document}